\documentclass[a4paper,11pt]{article}  
\usepackage{amsfonts}
\usepackage{tikz}
\usepackage{tikz,tikz-cd}
\usetikzlibrary{matrix,arrows}

\usepackage{latexsym}
\usepackage{amsthm}
\usepackage{amsmath}
\usepackage{graphicx}
\usepackage{multirow}
\usepackage{ulem}
\newtheorem{Le}{Lemma}

\newtheorem{Co}{Corollary}
\newtheorem{The}{Theorem}
\newtheorem{Pro}{Proposition}

\theoremstyle{definition}
\newtheorem{Exam}{Example}
\newtheorem{Property}{Property}
\newtheorem{Rem}{Remark}
\newtheorem{De}{Definition}

\newcommand{\Av}{\mathtt{Av}}
\newcommand{\Ss}{\mathfrak{S}}

\def\des{\mathsf{des}}
\def\inv{\mathrm{inv}}
\def\maj{\mathrm{maj}}
\def\mak{\mathrm{mak}}
\def\makl{\mathrm{makl}}
\def\fozep{\mathrm{foze'}}
\def\fozepp{\mathrm{foze''}}
\def\mad{\mathrm{mad}}
\def\sist{\mathrm{sist}}
\def\sistp{\mathrm{sist'}}
\def\sistpp{\mathrm{sist''}}
\def\bast{\mathrm{bast}}
\def\bastp{\mathrm{bast'}}
\def\bastpp{\mathrm{bast''}}
\def\foze{\mathrm{foze}}

\def\comp{\mathrm{c}}
\def\rev{\mathrm{r}}
\def\des{\mathrm{des}}
\def\Rlminl{\mathrm{Rlminl}}%
\def\Lrmax{\mathrm{Lrmax}}%
\def\Lrmaxl{\mathrm{Lrmaxl}}%
\def\Lrminl{\mathrm{Lrminl}}%
\def\Rlmaxl{\mathrm{Rlmaxl}}%
\def\Rlmin{\mathrm{Rlmin}}%

\def\Des{\mathrm{Des}}%
\def\Dbot{\mathrm{Dbot}}%
\def\Dtop{\mathrm{Dtop}}
\def\Asc{\mathrm{Asc}}%
\def\Atop{\mathrm{Atop}}%

\def\rev{\mathrm{r}}
\def\com{\mathrm{c}}

\title{The equidistribution of some Mahonian statistics 
over permutations avoiding a pattern of length three}
\author{Phan Thuan Do \footnote{Department of Computer Science, 
Hanoi University of Science and Technology, 01 Dai Co Viet, Hanoi, Vietnam,
email: thuandp@soict.hust.edu.vn} \and Thi Thu Huong Tran  \footnote{Vietnamese-German University, Le Lai street, Hoa Phu ward, Thu Dau Mot city, Binh Duong, Vietnam, email: huong.ttt@vgu.edu.vn} \and
Vincent Vajnovszki \footnote{LIB, Universit\'e de Bourgogne Franche-Comt\'e, B.P. 47 870, 21078 Dijon-Cedex, France, email: vvajnov@u-bourgogne.fr}}

\begin{document}

\maketitle

\begin{abstract}
We prove the equidistribution of several multistatistics over some classes of permutations avoiding a $3$-length pattern. 
We deduce the equidistribution, on the one hand of $\inv$ and $\fozepp$ statistics, and on the other hand that of $\maj$ and $\makl$ statistics, over these classes of pattern avoiding permutations. Here $\inv$ and $\maj$ are the celebrated Mahonian statistics,
$\fozepp$ is one of the statistics defined in terms of generalized patterns in the 2000 pioneering paper of Babson and Steingr{\'\i}msson, and $\makl$ is one of the statistics defined by Clarke, Steingr{\'\i}msson and Zeng in 1997.
These results solve several conjectures posed by Amini in~2018.
\end{abstract}

\section{Introduction}

Babson and Steingr{\'\i}msson \cite{BabSteim} introduced generalized permutation patterns and, based on linear combinations of functions counting occurrences of such patterns, they defined various statistics over permutations and identified many of them with well-known Mahonian statistics.
They made several conjectures regarding the Mahonian nature of new generalized pattern-based statistics. These have since been proved,  some of them refined and generalized in several different ways, see for instance \cite{FoataZeilberger,FHV,KV}.

In \cite{DDJSS} the notion of Wilf equivalence is generalized in the following way: for a statistic $\mathrm{st}$, the sets of classical patterns $\Pi$ and $\Sigma$ are $\mathrm{st}$-Wilf equivalent if $\mathrm{st}$ has the same distribution over the set of permutations avoiding each pattern in $\Pi$ and over the set of those avoiding each pattern in $\Sigma$. Among the results proved in \cite{DDJSS} are:
the inversion number, $\inv$, has the same distribution over $132$-avoiding (resp. $231$-avoiding) permutations and over $213$-avoiding (resp. $312$-avoiding) permutations; and the major index, $\maj$, has the same distribution over $132$-avoiding (resp. $213$-avoiding) permutations and over $231$-avoiding (resp. $312$-avoiding) permutations.

In \cite{Amini}, Amini exhaustively investigated quadruples $(\mathrm{st_1,st_2};\sigma,\tau)$ where $\mathrm{st_1}$ and $\mathrm{st_2}$ are Babson-Steingr{\'\i}msson's Mahonian statistics, and $\sigma$ and $\tau$ are $3$-length classical patterns. These quadruples must satisfy an equidistribution property, namely $\mathrm{st_1}$ over the set of $\sigma$-avoiding permutations is equidistributed with $\mathrm{st_2}$ over the set of $\tau$-avoiding permutations, or alternatively,
$$
\sum_{\pi \mbox{ \small avoids } \sigma} q^{\mathrm{st_1}\,\pi}\, t^{|\pi|}=
\sum_{\pi \mbox{ \small avoids } \tau} q^{\mathrm{st_2}\,\pi}\, t^{|\pi|},
$$
where $|\pi|$ is the length of the permutation $\pi$.

For instance, with this terminology, an aforementioned result in \cite{DDJSS} is that $(\inv, \inv; 132,213)$ satisfies the equidistribution property. 
In \cite{Amini} the equidistribution property is proved for many such quadruples, and for others it is still conjectured.
Chen \cite{JoannaChen} based on a preprint version of \cite{Amini} settled the conjectures corresponding to $\mathrm{st_1}=\maj$ and  $\mathrm{st_2}=\bast$. 
Note that Chen used the notation \textsc{stat}
adopted from \cite{BabSteim} (and used in other papers \cite{FHV,KV}) instead of $\bast$ as in \cite{Amini}, and through this paper we adhere to Amini's \cite{Amini} notations for Babson-Steingr{\'\i}msson's Mahonian statistics.

Other statistic in \cite{BabSteim} is $\fozepp$, whose Mahonian nature (up to reverse operation) is proved in \cite[Theorem 2]{FoataZeilberger}
and $\makl$ introduced in \cite{CSZ} in the context of Mahonian statistics on words.
In the present paper we solve the conjectures in \cite{Amini} corresponding to $\mathrm{st_1}=\fozepp$ and  $\mathrm{st_2}=\inv$,
and to $\mathrm{st_1}=\makl$ and  $\mathrm{st_2}=\maj$.
%
%
In both cases we are able to find more refined equidistributions, and other conjectures in 
\cite{Amini} are solved by combining these results with other known equidistributions.

\section{Notations and definitions}

We denote by $\Ss_n$ the set of $n$-length permutations, and $\Ss=\cup_{n\geq 0}\Ss_n$.
Let $\sigma\in \Ss_k$ and $\pi=\pi_1\pi_2\dots \pi_n\in \Ss_n$, $1\leq k \leq n$, be two permutations. One says that $\pi$ contains $\sigma$ if $\pi$ contains a subsequence $\pi_{i_1}\pi_{i_2}\dots\pi_{i_k}$, $i_1<i_2<\dots <i_k$, order isomorphic to $\sigma$; otherwise  one says that $\pi$ avoids $\sigma$, or $\pi$ is $\sigma$-avoiding. In this context $\sigma$ is called a (classical) {\it pattern}.
For a pattern $\sigma$, $\Av_n(\sigma)$ denotes the set of $\sigma$-avoiding permutations in $\Ss_n$, and $\Av(\sigma)=\cup_{n\geq 0}\Av_n(\sigma)$.
{\it Generalized patterns} have been introduced in \cite{BabSteim} and they were extensively studied since then. See for instance Chapter 7 in \cite{KitaevBook} for a comprehensive description of results on these patterns. 
A particular case of generalized patterns is that where two adjacent letters may be underlined, which means that the corresponding letters in the permutation must be adjacent. 
For example, the pattern $\underline{31}2$ occurs in the permutation $361524$ four times, namely, as the subsequences 
 $615$, $612$, $614$ and $524$. Note that the subsequences $312$ and $624$ are not occurrences of $\underline{31}2$ because their first two letters are not adjacent in the permutation. 

\medskip

A combinatorial {\it statistic} over $\Ss_n$ is simply a function defined on $\Ss_n$.
Here we consider statistics whose codomains are integers, sets of integers 
or sets of pairs of integers. The distribution of an integer valued statistic $\mathrm{st}$ over the set $S\subset \Ss_n$ is given by the coefficients of the generating function $\sum_{\sigma\in S}q^{\mathrm{st}\,\sigma}$.
The overall number of occurrences of the patterns $\sigma_1,\sigma_2,\dots,\sigma_{\ell}$ in the permutation $\pi$ becomes an integer valued statistic, denoted $(\sigma_1+\sigma_2+\dots+\sigma_{\ell})\,\pi$. 
The {\it descent number} statistic is defined as $\des\,\pi=(\underline{21})\,\pi$.
Another classical example of integer valued statistic is $\inv$, the {\it inversion number}: an inversion in a permutation $\pi$ is a pair $(i,j)$ with $i<j$ and $\pi_i>\pi_j$, and $\inv\,\pi$ is the number of inversions of $\pi$. Alternatively, in terms of functions counting occurrences of generalized patterns, it is easily seen that
$$\inv\, \pi=\left(\underline{23}1+\underline{31}2+\underline{32}1+\underline{21}\right)\pi,$$
and any integer valued statistic over $\Ss_n$ which has the same distribution as $\inv$ is called Mahonian. 
The first statistic proved to be equidistributed with $\inv$ (and thus Mahonian) is $\maj$ \cite{MacMahon}, defined as $\maj\,\pi=\sum_{\pi_i>\pi_{i+1}}i$, or alternatively, by counting for each descent the number of entries in $\pi$ to its left,
$$\maj\,\pi=\left(1\underline{32}+2\underline{31}+3\underline{21}+\underline{21}\right)\pi.$$
Babson-Steingr{\'\i}msson's \cite{BabSteim} statistics considered in \cite{Amini} are (with the notations in \cite{Amini}):

\begin{tabular}{l c l}
$\mad\,\pi=\left(2\underline{31}+2\underline{31}+\underline{31}2+\underline{21}\right)\pi$
& & $\foze\,\pi=\left(\underline{21}3+3\underline{21}+\underline{13}2+\underline{21}\right)\pi$\\
$\mak\,\pi=\left(1\underline{32}+\underline{31}2+\underline{32}1+\underline{21}\right)\pi$
& & $\fozep\,\pi=\left(1\underline{32}+2\underline{31}+2\underline{31}+\underline{21}\right)\pi$\\
$\makl\, \pi=\left(1\underline{32}+2\underline{31}+\underline{32}1+\underline{21}\right)\pi$
& & $\fozepp\,\pi=\left(\underline{23}1+\underline{31}2+\underline{31}2+\underline{21}\right)\pi$\\
$\bast\,\pi=\left(\underline{13}2+\underline{21}3+\underline{32}1+\underline{21}\right)\pi$
& & $\sist\,\pi=\left(\underline{13}2+\underline{13}2+2\underline{13}+\underline{21}\right)\pi$\\
$\bastp\,\pi=\left(\underline{13}2+\underline{31}2+\underline{32}1+\underline{21}\right)\pi$
& & $\sistp\,\pi=\left(\underline{13}2+\underline{13}2+2\underline{31}+\underline{21}\right)\pi$\\
$\bastpp\,\pi=\left(1\underline{32}+3\underline{12}+3\underline{21}+\underline{21}\right)\pi$
& & $\sistpp\,\pi=\left(\underline{13}2+2\underline{31}+2\underline{31}+\underline{21}\right)\pi$,
\end{tabular}

\noindent
see \cite{CSZ} for alternative definitions of some of these statistics.

\medskip

For a permutation $\pi=\pi_1\pi_2\dots \pi_n$, its {\it reverse} $\rev(\pi)$ is the permutation $\pi_n\pi_{n-1}\dots \pi_1$,
and its {\it complement} $\com(\pi)$ is the permutation $(n-\pi_1+1)(n-\pi_2+1)\dots (n-\pi_n+1)$.

A {\it left-to-right maximum} of $\pi$ is a pair $(i,\pi_i)$ with $\pi_i>\pi_j$ for all $j<i$; $i$ is the {\it position} and $\pi_i$ is the {\it letter} of the left-to-right maximum. We denote by $\Lrmax\,\pi$ the set of left-to-right maxima of $\pi$ and by 
$\Lrmaxl\,\pi$ the set of letters of left-to-right maxima of $\pi$. 
We define similarly a {\it left-to-right minimum}, {\it right-to-left maximum} and {\it right-to-left minimum} of $\pi$, and 
$\Lrminl$, $\Rlmaxl$, $\Rlmin$ and $\Rlminl$ have obvious meaning.

In any occurrence $\pi_i\pi_{i+1}$ of the pattern $\underline{21}$ in the permutation $\pi$ (thus, $\pi_i>\pi_{i+1}$), the position $i$ is called {\it descent}, the value 
$\pi_i$ is called {\it descent top} and $\pi_{i+1}$ {\it descent bottom}; $\Des\,\pi$, $\Dtop\,\pi$ and $\Dbot\,\pi$ denotes, respectively, the set of descents, descent tops and descent bottoms of $\pi$. Similarly, if $\pi_i<\pi_{i+1}$, then the position $i$ is called {\it ascent} and the value $\pi_{i+1}$ is called {\it ascent top}; $\Asc\,\pi$ and $\Atop\,\pi$ denotes, respectively, the set of ascents and of ascent tops of $\pi$.

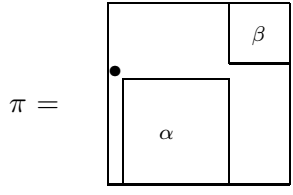
\begin{figure}
\begin{center}
\begin{tabular}{rl}
\begin{tabular}{c}
$\pi$ = \\
\\
\\
\\
\\
\end{tabular}
&
\unitlength=2mm
\begin{picture}(12,10)
\put(0,0){\line(1,0){12}}
\put(0,12){\line(1,0){12}}
\put(0,0){\line(0,1){12}}
\put(12,0){\line(0,0){12}}
\put(1,0){\line(0,1){7}}
\put(1,7){\line(1,0){7}}
\put(8,0){\line(0,1){7}}
\put(8,8){\line(1,0){4}}
\put(8,8){\line(0,1){4}}
\put(0.5,7.5){\circle*{0.7}} 

\put(9.5,9.5){$\scriptstyle \beta$}
\put(3.4,3.0){$\scriptstyle \alpha$}
\end{picture} 
\end{tabular}
\end{center}
\caption{The decomposition $\pi=(1\ominus\alpha)\oplus\beta$ of a non-empty $231$-avoiding permutation $\pi$.
\label{dec_av_231}}
\end{figure}

For a permutation $\pi$, $|\pi|$ is its length (and so, $\pi\in \Ss_{|\pi|}$),
and for two permutations $\alpha$ and~$\beta$, their {\it direct sum}, denoted $\alpha\oplus\beta$, 
is the permutation $\pi$ of length $|\alpha|+|\beta|$ with
$$
\pi_i=
\left\{ \begin {array}{ccl}
\alpha_i & {\rm if} & 1\leq i\leq |\alpha| \\
\beta_{i-|\alpha|}+|\alpha| & {\rm if} & |\alpha|+1\leq i\leq |\alpha|+|\beta|,
\end {array}
\right.
$$
and their {\it skew sum}, denoted $\alpha\ominus\beta$, is the permutation $\pi$ of length $|\alpha|+|\beta|$ with
$$
\pi_i=
\left\{ \begin {array}{ccl}
\alpha_i+|\beta| & {\rm if} & 1\leq i\leq |\alpha| \\
\beta_{i-|\alpha|} & {\rm if} & |\alpha|+1\leq i\leq |\alpha|+|\beta|.
\end {array}
\right.
$$
The following lemma regarding the structure of $\Av(231)$ is part of the folklore of pattern avoidance (see for instance \cite{KitaevBook}).
\begin{Le}
Any non-empty permutation $\pi$ is $231$-avoiding if and only if it can be written in a unique way as $\pi=(1\ominus\alpha)\oplus\beta$, for some $231$-avoiding permutations $\alpha$ and $\beta$.
\end{Le}
See the Figure \ref{dec_av_231} for the diagram representation of such a permutation $\pi$.

\section{Equidistributions involving $\fozepp$ and $\inv$}
\label{Sec3}
The main results of this section are Theorems \ref{inv_231} and \ref{inv_321_312}. Some of their consequences  are summarized in Table 
\ref{Tab_Summ}.

\subsection{The equidistribution of $\fozepp$ and $\inv$ over $\Av_n(231)$}
\label{Av_231}

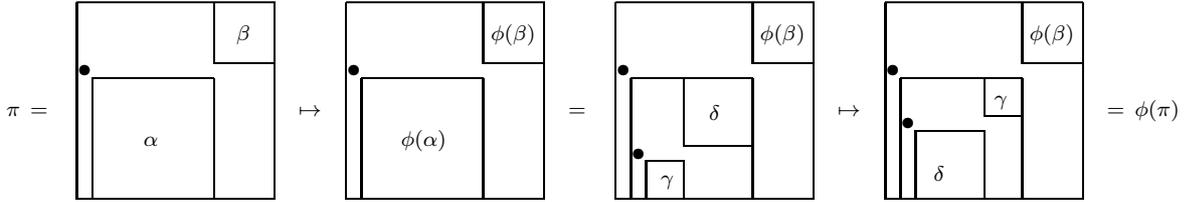
\begin{figure}[h]

\begin{tabular}{ccccccccc}
\unitlength=2mm
\begin{picture}(0.5,5)\put(-2.0,5.5){$\scriptstyle \pi\ =\ $}
\end{picture}
&
\unitlength=2mm
\begin{picture}(13,13)
\put(0,0){\line(1,0){13}}
\put(0,13){\line(1,0){13}}
\put(0,0){\line(0,1){13}}
\put(13,0){\line(0,0){13}}
\put(1,0){\line(0,1){8}}
\put(1,8){\line(1,0){8}}
\put(9,0){\line(0,1){8}}
\put(9,9){\line(1,0){4}}
\put(9,9){\line(0,1){4}}
\put(0.5,8.5){\circle*{0.7}} 

\put(10.5,10.5){$\scriptstyle \beta$}
\put(4.4,3.5){$\scriptstyle \alpha$}
\end{picture} 
&
\unitlength=2mm
\begin{picture}(0.5,0)\put(-0.5,5.5){$\scriptstyle \mapsto$}
\end{picture} 
&
\unitlength=2mm
\begin{picture}(13,13)
\put(0,0){\line(1,0){13}}
\put(0,13){\line(1,0){13}}
\put(0,0){\line(0,1){13}}
\put(13,0){\line(0,0){13}}
\put(1,0){\line(0,1){8}}
\put(1,8){\line(1,0){8}}
\put(9,0){\line(0,1){8}}
\put(9,9){\line(1,0){4}}
\put(9,9){\line(0,1){4}}
\put(0.5,8.5){\circle*{0.7}} 

\put(9.5,10.5){$\scriptstyle \phi(\beta)$}
\put(3.6,3.5){$\scriptstyle \phi(\alpha)$}
\end{picture} 
&
\unitlength=2mm
\begin{picture}(0.5,0)\put(-0.5,5.5){$\scriptstyle =$}
\end{picture}
&
\unitlength=2mm
\begin{picture}(13,13)
\put(0,0){\line(1,0){13}}
\put(0,13){\line(1,0){13}}
\put(0,0){\line(0,1){13}}
\put(13,0){\line(0,0){13}}
\put(1,0){\line(0,1){8}}
\put(1,8){\line(1,0){8}}
\put(9,0){\line(0,1){8}}
\put(1.5,3.0){\circle*{0.7}} 
\put(2,0){\line(0,1){2.5}}
\put(4.5,0){\line(0,1){2.5}}
\put(2,2.5){\line(1,0){2.5}}
\put(4.5,3.5){\line(1,0){4.5}}
\put(4.5,3.5){\line(0,1){4.5}}
\put(9,9){\line(1,0){4}}
\put(9,9){\line(0,1){4}}
\put(0.5,8.5){\circle*{0.7}} 

\put(9.5,10.5){$\scriptstyle \phi(\beta)$}
\put(6.125,5.25){$\scriptstyle \delta$}
\put(2.95,0.95){$\scriptstyle \gamma$}
\end{picture}
&\unitlength=2mm
\begin{picture}(0.5,0)\put(-0.5,5.5){$\scriptstyle \mapsto$}
\end{picture} 
&
\unitlength=2mm
\begin{picture}(13,13)
\put(0,0){\line(1,0){13}}
\put(0,13){\line(1,0){13}}
\put(0,0){\line(0,1){13}}
\put(13,0){\line(0,0){13}}
\put(1,0){\line(0,1){8}}
\put(1,8){\line(1,0){8}}
\put(9,0){\line(0,1){8}}
\put(1.5,5){\circle*{0.7}} 
\put(2,0){\line(0,1){4.5}}
\put(6.5,0){\line(0,1){4.5}}
\put(2.0,4.5){\line(1,0){4.5}}

\put(6.5,5.5){\line(1,0){2.5}}
\put(6.5,5.5){\line(0,1){2.5}}
\put(9,9){\line(1,0){4}}
\put(9,9){\line(0,1){4}}
\put(0.5,8.5){\circle*{0.7}} 

\put(9.5,10.5){$\scriptstyle \phi(\beta)$}
\put(7.2,6.2){$\scriptstyle \gamma$}
\put(3.2,1.2){$\scriptstyle \delta$}
\end{picture} 
&
\unitlength=2mm
\begin{picture}(0.5,0)\put(-0.5,5.5){$\scriptstyle =\ \phi(\pi)$}
\end{picture}
\end{tabular}
\caption{\label{decomposition}
The construction of the image of $\pi=(1\ominus \alpha)\oplus\beta$ through  $\phi$
when $\alpha$ is not empty.
}
\end{figure}

We define recursively the map $\phi$ on $\Av(231)$ as:
\begin{itemize}
\item[$(i)$] if $\pi=\lambda$, then $\phi(\pi)=\lambda$, where $\lambda$ is the empty permutation,
\item[$(ii)$] if $\pi=1\oplus\beta$, then $\phi(\pi)=1\oplus\phi(\beta)$, and
\item[$(iii)$] if $\pi=(1\ominus \alpha)\oplus\beta$ with $\alpha\neq \lambda$, then
$\phi(\pi)=(1\ominus((1\ominus\delta)\oplus\gamma))\oplus\phi(\beta)$, where $\gamma$ and $\delta$ are such that 
$\phi(\alpha)=(1\ominus \gamma)\oplus\delta$.
\end{itemize} 
\noindent 
See Figure \ref{decomposition} for the diagram representation  of $\phi(\pi)$ in the case $(iii)$.

\begin{Rem}
By construction, $\phi$ is length preserving and $\phi(\pi)\in \Av(231)$ whenever $\pi\in \Av(231)$. Moreover, $\phi$ keeps the first entry and the statistic $\Lrmax$ of a permutation in its preimage and image.
\label{Rem1}
\end{Rem}

\noindent
Our map  $\phi$ is defined on $\Av(231)$, and by a slight abuse of notation we denote also by $\phi$ its restriction to $\Av_n(231)$. 
By induction on $n$ it can be seen that $\phi$ on $\Av_n(231)$ 
is invertible, so $\phi$ is a bijection (but in general $\phi$ is not an involution). 

In the proof of the next theorem we need the following remark where the equality involving 
$\fozepp$ statistic is based on the observation that if $p$ is one of the patterns 
$\underline{23}1$, $\underline{31}2$ or $\underline{21}$, then $(p)\,\alpha\oplus\beta=(p)\,\alpha+(p)\,\beta$.

\begin{Rem}
For any two permutations $\alpha$ and $\beta$ we have that $\inv\,\alpha\oplus\beta=\inv\,\alpha+\inv\,\beta$ and $\fozepp\,\alpha\oplus\beta=\fozepp\,\alpha+\fozepp\,\beta$.
\label{Rem2}
\end{Rem}

\begin{The}
\label{inv_231}
The bistatistics $(\fozepp,\Lrmax)$ and $(\inv,\Lrmax)$ have the same distribution over $\Av_n(231)$.
\end{The}
\proof
We show by induction on $n$ that $(\inv,\Lrmax)\,\phi(\pi)=(\fozepp,\Lrmax)\,\pi$ for any $\pi\in\Av_n(231)$.
By Remark \ref{Rem1} it is sufficient to show that $\inv\,\phi(\pi)=\fozepp\,\pi$, and by Remark~\ref{Rem2} it is  sufficient  to prove that if $\tau=1\ominus \alpha$, with $\alpha$ non-empty, then $\inv\,\phi(\tau)=\fozepp\,\tau$. Note that $\tau$ is the first term of $\pi$ in point $(iii)$ defining $\phi$, and thus $\phi(\tau)=1\ominus((1\ominus\delta)\oplus\gamma)$, where $\gamma$ and $\delta$ are such that $\phi(\alpha)=(1\ominus \gamma)\oplus\delta$. We refer the reader to Figure~\ref{decomposition} reading it from right to left, disregarding $\beta$ and $\phi(\beta)$. We have 
\begin{eqnarray*}
\inv\,\phi(\tau) & = & \inv\,1\ominus((1\ominus\delta)\oplus\gamma) \\
& = & \inv\,\delta+ \inv\,\gamma+2|\delta|+|\gamma|+1\\
& = & \inv\,(1\ominus\gamma)\oplus\delta+2|\delta|+1,\\
& = & \inv\,\phi(\alpha)+2|\delta|+1.
\end{eqnarray*}

\noindent
Since $\phi$ preserves the first element of a permutation, it follows that if $\phi(\alpha)=(1\ominus \gamma) \oplus \delta $, then $\alpha=(1\ominus \gamma') \oplus \delta'$ with $|\delta|=|\delta'|$ (and $|\gamma|=|\gamma'|$). 
With these notations, by the induction hypothesis we have 
\begin{eqnarray*}
\inv\,\phi(\alpha)+2|\delta|+1 & = & \fozepp\,\alpha+2|\delta'|+1.
\end{eqnarray*}
For any permutation $\alpha$ we have 
$(\underline{23}1)\,1\ominus \alpha=(\underline{23}1)\,\alpha$, and
$(\underline{21})\,1\ominus \alpha=(\underline{21})\,\alpha+1$.
In addition, since $\alpha=(1\ominus \gamma') \oplus \delta'$ it follows that 
$(\underline{31}2)\,1\ominus \alpha=(\underline{31}2)\,\alpha+|\delta'|$. 
Finally, since $\fozepp$ is a `linear combination' of $(\underline{23}1)$, $(\underline{31}2)$ and $(\underline{21})$
we have 
\begin{eqnarray*}
\fozepp\,\alpha+2|\delta'|+1 & = & \fozepp\,1\ominus \alpha\\
& = & \fozepp\,\tau,
\end{eqnarray*}
and the statement holds.
\endproof

\begin{Exam}
If $\pi=321$, then $\phi(\pi)=312$, and if $\pi=321654$, then $\phi(\pi)=312645$. In the last case $\fozepp\,\pi=\inv\,\phi(\pi)=4$ and $\Lrmax\,\pi=\Lrmax\,\phi(\pi)=\{(1,3),(4,6)\}$.
\end{Exam}

%

\subsection{The equidistribution of $\fozepp$ over $\Av_n(312)$ and $\inv$ over $\Av_n(321)$}
\label{Av_312_321}

We recall the definition of the well-known bijection $\psi:\Av_n(312)\to \Av_n(321)$ due to Simion and Schmidt \cite{SS}.
The map $\pi\overset{\psi}{\mapsto}\sigma$ is defined as: keep the left-to-right maxima of $\pi$ fixed, and write all the other entries in increasing order. Clearly, $\sigma$ is $321$-avoiding as it is the union of two increasing subsequences, one of which is the sequence of left-to-right maxima and the other is the increasing sequence of the remaining entries. In \cite{SS} it is shown that $\psi$ is a bijection, see also \cite[Lemma 4.3]{BonaBook} (up to complement operation) and Figure \ref{SS_transform} for an example.

\begin{The}
\label{inv_321_312}
The bistatistic 
$(\fozepp,\Lrmax)$ over $\Av_n(312)$ has the same distribution as
$(\inv,\Lrmax)$ over $\Av_n(321)$.
\end{The}
\proof
Bijection  $\psi$ preserves $\Lrmax$ statistic, thus it is sufficient to prove that $\fozepp\ \pi=\inv\ \psi(\pi)$ for $\pi\in \Av_n(312)$; and since $\pi$ is $312$-avoiding it is equivalent to prove that $\left(\underline{23}1+\underline{21}\right)\pi=\inv\ \psi(\pi)$.
We proceed by induction on $n$.
Let $\pi=\pi_1\pi_2\dots\pi_n$ be a permutation in $\Av_n(312)$, $n\geq 2$. We distinguish two cases: (i) $\pi_2>\pi_1$, and (ii) otherwise.

(i) Let $\pi'$ be the permutation in $\Av_{n-1}(312)$ obtained from $\pi$ after deleting its first entry and re-scaling the remaining entries to a permutation, and let $\sigma'=\psi(\pi')$. Clearly, $\sigma=\psi(\pi)$ is obtained from $\sigma'$ by inserting $\pi_1$ in front of it, and adding $1$
to all entries in $\sigma'$ larger than or equal to $\pi_1$. We have
\begin{itemize}
\item $\left(\underline{23}1+\underline{21}\right)\pi=\left(\underline{23}1+\underline{21}\right)\pi'+u$, where $u$ is the number of occurrences of $\underline{23}1$ in $\pi$
where the role of $2$ is played by $\pi_1$, or equivalently, the number of entries in $\pi$  less than $\pi_1$, and
\item $\inv\ \sigma=\inv\ \sigma'+v$, where $v$ is the number of inversions in $\sigma$  involving $\sigma_1=\pi_1$.
\end{itemize}
Clearly, $u=v$, and by the induction hypothesis $\left(\underline{23}1+\underline{21}\right)\pi'=\inv\ \sigma'$, and the statement holds.

(ii) Since $\pi$ avoids $312$, it follows that $\pi_1=\pi_2+1$.
Let $\pi'$ be the permutation obtained from $\pi$ by deleting $\pi_2$ and re-scaling the remaining entries to a permutation, and let $\sigma'=\psi(\pi')$. The permutation $\sigma=\psi(\pi)$ is obtained from $\sigma'$ by adding $1$ to each of its entries, then 
by inserting $1$ between the entries $\sigma'_1$ and $\sigma'_2$.
We have
\begin{itemize}
\item $\left(\underline{23}1+\underline{21}\right)\pi=\left(\underline{23}1+\underline{21}\right)\pi'+1$ because $\pi$ and $\pi'$ have the same numbers of occurrences of $\underline{23}1$ but $\pi$ has one more descent than $\pi'$, namely $\pi_1\pi_2$,
and
\item $\inv\ \sigma=\inv\ \sigma'+1$.
\end{itemize}
Again, by the induction hypothesis, the statement holds.
\endproof

\begin{figure}
\begin{center}

\begin{tabular}{rcl}

\unitlength=4mm
\begin{picture}(8,8)
\put(0,0){\line(1,0){8}}
  
\put(0,8){\line(1,0){8}}
\put(0,0){\line(0,1){8}}
%
\put(8,0){\line(0,0){8}}
\put(1,0){\line(0,1){8}}
\put(2,0){\line(0,1){8}}
\put(3,0){\line(0,1){8}}
\put(4,0){\line(0,1){8}}
\put(5,0){\line(0,1){8}}
\put(6,0){\line(0,1){8}}
\put(7,0){\line(0,1){8}}

\put(0,1){\line(1,0){8}}
\put(0,2){\line(1,0){8}}
\put(0,3){\line(1,0){8}}
\put(0,4){\line(1,0){8}}
\put(0,5){\line(1,0){8}}
\put(0,6){\line(1,0){8}}
\put(0,7){\line(1,0){8}}

\put(0.5,2.5){\circle*{0.4}} 
\put(1.5,1.5){\circle{0.4}}
\put(2.5,5.5){\circle*{0.4}}
\put(3.5,4.5){\circle{0.4}}
\put(4.5,7.5){\circle*{0.4}}
\put(5.5,6.5){\circle{0.4}}
\put(6.5,3.5){\circle{0.4}}
\put(7.5,0.5){\circle{0.4}}
\put(0.3,3.1){$\scriptstyle 3$}
\put(2.3,6.1){$\scriptstyle 6$}
\put(3.4,7.3){$\scriptstyle 8$}
\put(1.3,0.3){$\scriptstyle 2$}
\put(4.3,4.3){$\scriptstyle 5$}
\put(5.3,5.3){$\scriptstyle 7$}
\put(6.3,2.3){$\scriptstyle 4$}
\put(6.45,0.3){$\scriptstyle 1$}
\end{picture}
& 
\begin{tabular}{c}
$\overset{\psi}{\longmapsto}$\\\
\\
\\
\\
\\
\\
\\
\end{tabular}
&
\unitlength=4mm
\begin{picture}(8,8)
\put(0,0){\line(1,0){8}}
%
\put(0,8){\line(1,0){8}}
\put(0,0){\line(0,1){8}}

\put(8,0){\line(0,0){8}}

\put(1,0){\line(0,1){8}}
\put(2,0){\line(0,1){8}}
\put(3,0){\line(0,1){8}}
\put(4,0){\line(0,1){8}}
\put(5,0){\line(0,1){8}}
\put(6,0){\line(0,1){8}}
\put(7,0){\line(0,1){8}}

\put(0,1){\line(1,0){8}}
\put(0,2){\line(1,0){8}}
\put(0,3){\line(1,0){8}}
\put(0,4){\line(1,0){8}}
\put(0,5){\line(1,0){8}}
\put(0,6){\line(1,0){8}}
\put(0,7){\line(1,0){8}}

\put(0.5,2.5){\circle*{0.4}} 
\put(1.5,0.5){\circle{0.4}} 
\put(2.5,5.5){\circle*{0.4}} 
\put(3.5,1.5){\circle{0.4}} 
\put(4.5,7.5){\circle*{0.4}} 
\put(5.5,3.5){\circle{0.4}} 
\put(6.5,4.5){\circle{0.4}} 
\put(7.5,6.5){\circle{0.4}} 
\put(0.3,3.1){$\scriptstyle 3$}
\put(2.3,6.1){$\scriptstyle 6$}
\put(3.4,7.3){$\scriptstyle 8$}
\put(0.45,0.3){$\scriptstyle 1$}
\put(3.3,0.3){$\scriptstyle 2$}
\put(5.3,2.3){$\scriptstyle 4$}
\put(7.3,4.3){$\scriptstyle 5$}
\put(7.3,5.3){$\scriptstyle 7$}
\end{picture} 
\end{tabular}

\vspace{-1.5cm}
\end{center}
\caption{\label{SS_transform}
The bijection $\psi$ transforms the permutation $\pi=32658741\in\Av(312)$ into $\sigma=31628457\in\Av(321)$. Left-to-right maxima are black.}
\end{figure}
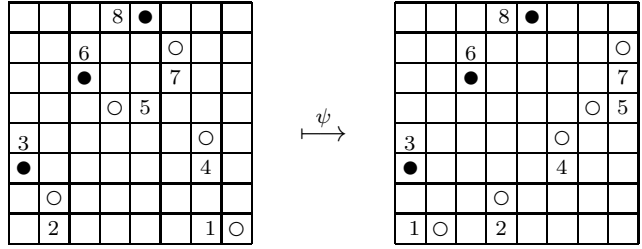
\begin{Exam}
If $\pi$ and $\sigma$ are the permutations in Figure \ref{SS_transform}, then $\fozepp\,\pi=\inv\,\sigma=8$ and $\Lrmax\,\pi=\Lrmax\,\sigma=\{(1,3),(3,6),(5,8)\}$.
\end{Exam}

\subsection{Some consequences}

We conclude this section  with a couple of consequences of Theorems \ref{inv_231} and \ref{inv_321_312}.

Since $\inv$ has the same distribution over $\Av_n(231)$ and $\Av_n(312)$, see \cite{DDJSS}, we have the next corollary.

\begin{Co}
\label{fozepp_plus}
The statistic $\fozepp$ over $\Av_n(231)$ has the same distribution as $\inv$ over $\Av_n(312)$.
\end{Co}

\begin{Co}
\label{mad}
The statistic $\fozepp$ over $\Av_n(312)$ (resp. over $\Av_n(231)$) has the same distribution as $\mad$ over $\Av_n(231)$ (resp. over $\Av_n(312)$).
\end{Co}
\proof
By Theorem \ref{inv_321_312}, $\fozepp$ over $\Av(312)$ has the same distribution as
$\inv$ over $\Av_n(321)$, which in turn by \cite[Corollary 24]{Amini},
has the same distribution as $\mad$ over $\Av_n(231)$.

\noindent
By Theorem \ref{inv_231}, $\fozepp$ over $\Av_n(231)$ has the same distribution as 
$\inv$ over $\Av_n(231)$, or equivalently applying $\rev\circ \comp$ operation, as 
$\inv$ over $\Av_n(312)$, which in turn has the same distribution as $\mad$ over $\Av_n(312)$, see \cite{KMPW}.
(Recall that $\rev\circ \comp$ is the reverse-complement operation, and it preserves $\inv$ statistic.)
\endproof

It is worth mentioning that the bistatistic $(\fozepp,\mad)$ over $\Av_n(312)$ does not have the same distribution as $(\fozepp,\mad)$ nor as $(\mad,\fozepp)$ over $\Av_n(231)$.

\medskip

Combining Theorems \ref{inv_231} and \ref{inv_321_312} with equidistributions in \cite{Amini} and reasoning as above
we have the next corollary.

\begin{Co}
\label{other} The statistic $\fozepp$ 
\begin{enumerate} 
\item over $\Av_n(231)$ has the same distribution as $\fozep$ over $\Av_n(132)$,
\item over $\Av_n(231)$ (resp. over $\Av_n(312)$) has the same distribution as $\sist$ over $\Av_n(213)$ (resp. over $\Av_n(132)$),
\item over $\Av_n(312)$ (resp. over $\Av_n(231)$) has the same distribution as $\sistp$ over $\Av_n(132)$ (resp. over $\Av_n(231)$),
\item over $\Av_n(231)$ (resp. over $\Av_n(312)$) has the same distribution as $\sistpp$ over $\Av_n(132)$ (resp. over $\Av_n(231)$).
\end{enumerate}
\end{Co}


\begin{table}
\begin{center}
\begin{tabular}{|l|cc|}
  \hline
  \multirow{3}{*}{$\inv,\fozepp$} & $231,231$ & Th. \ref{inv_231}\\
    & $321,312$  & Th. \ref{inv_321_312}\\ 
    & 312,231   & Cor. \ref{fozepp_plus}   \\ \hline
  \multirow{2}{*}{$\mad,\fozepp$} & $231,312$ & \multirow{2}{*}{Cor. \ref{mad}}\\
    & $312,231$ & \\
 \hline
$\fozep,\fozepp$ & $132,231$ & Cor. \ref{other}.1\\
\hline
  \multirow{2}{*}{$\sist,\fozepp$} & $213,231$ & \multirow{2}{*}{Cor. \ref{other}.2}\\
    & $132,312$ & \\
\hline
  \multirow{2}{*}{$\sistp,\fozepp$} & $132,312$ & \multirow{2}{*}{Cor. \ref{other}.3}\\
    & $231,231$ & \\
 \hline
  \multirow{2}{*}{$\sistpp,\fozepp$} & $132,231$ & \multirow{2}{*}{Cor. \ref{other}.4}\\
    & $231,312$ & \\
 \hline
\end{tabular}
\end{center}
\caption{
\label{Tab_Summ}
Equidistribution of $\fozepp$ and other Babson-Steingr{\'\i}msson's Mahonian statistic conjectured in \cite{Amini} and proved in Section \ref{Sec3}.
It must be read, for instance, as: $\inv$ and $\fozepp$ have the same distribution over $\Av_n(231)$; and $\inv$ over $\Av_n(321)$ has the same distribution as $\fozepp$ over $\Av_n(312)$. Computer experiments
have shown that these are, up to trivial transformations,  the only such equidistributions involving $\fozepp$.
}
\end{table}

\section{Equidistributions involving $\maj$ and $\makl$}

In this section, we turn our attention to the statistics $\maj$ and $\makl$ to prove similar equidistribution results. The main results of this section are Theorem \ref{The_maj_makl}, and Propositions \ref{main_231} and \ref{Av_312}. Some of their consequences  are summarized in Table 
\ref{Tab_Summ_makl}.
\label{S3}
\subsection{The equidistribution of $\maj$ and $\makl$ over $\Av_n(231)$}
\label{first_312}

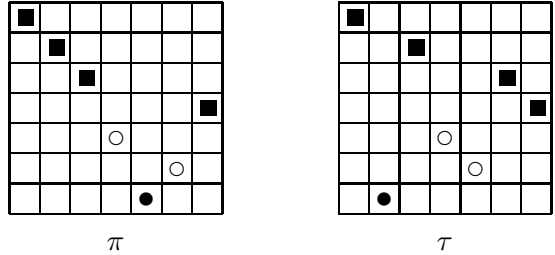
\begin{figure}[h]
\begin{center}  
\begin{tabular}{ccc}
\unitlength=4mm
\begin{picture}(7,7)
\put(0.,0.){\line(1,0){7}}
\put(0.,1.){\line(1,0){7}}
\put(0.,2.){\line(1,0){7}}
\put(0.,3.){\line(1,0){7}}
\put(0.,4.){\line(1,0){7}}
\put(0.,5.){\line(1,0){7}}
\put(0.,6.){\line(1,0){7}}
\put(0.,7.){\line(1,0){7}}
\put(0.,0.){\line(0,1){7}}
\put(1.,0.){\line(0,1){7}}
\put(2.,0.){\line(0,1){7}}
\put(3.,0.){\line(0,1){7}}
\put(4.,0.){\line(0,1){7}}
\put(5.,0.){\line(0,1){7}}
\put(6.,0.){\line(0,1){7}}
\put(7.,0.){\line(0,1){7}}
\put(0.3,6.25){\rule{0.2cm}{0.2cm}}
\put(1.3,5.25){\rule{0.2cm}{0.2cm}}
\put(2.3,4.25){\rule{0.2cm}{0.2cm}}
\put(6.3,3.25){\rule{0.2cm}{0.2cm}}

\put(3.5,2.5){\circle{0.4}}
\put(5.5,1.5){\circle{0.4}}

\put(4.5,0.5){\circle*{0.4}} 
\end{picture}
& 
\phantom{toto}
&
\unitlength=4mm
\begin{picture}(7,7)
\put(0.,0.){\line(1,0){7}}
\put(0.,1.){\line(1,0){7}}
\put(0.,2.){\line(1,0){7}}
\put(0.,3.){\line(1,0){7}}
\put(0.,4.){\line(1,0){7}}
\put(0.,5.){\line(1,0){7}}
\put(0.,6.){\line(1,0){7}}
\put(0.,7.){\line(1,0){7}}
\put(0.,0.){\line(0,1){7}}
\put(1.,0.){\line(0,1){7}}
\put(2.,0.){\line(0,1){7}}
\put(3.,0.){\line(0,1){7}}
\put(4.,0.){\line(0,1){7}}
\put(5.,0.){\line(0,1){7}}
\put(6.,0.){\line(0,1){7}}
\put(7.,0.){\line(0,1){7}}
\put(0.3,6.25){\rule{0.2cm}{0.2cm}}
\put(2.3,5.25){\rule{0.2cm}{0.2cm}}
\put(5.3,4.25){\rule{0.2cm}{0.2cm}}
\put(6.3,3.25){\rule{0.2cm}{0.2cm}}

\put(4.5,1.5){\circle{0.4}}
\put(3.5,2.5){\circle{0.4}}

\put(1.5,0.5){\circle*{0.4}} 
\end{picture}
\\
$\pi$ & & $\tau$
\end{tabular}
\end{center}
\caption{
\label{fig_theta}
The $231$-avoiding permutations $\pi=7653124$ and $\tau=\theta'(\pi)=7163254$.
The three i.d.r.'s of $\pi$ are $\{1,2,3,7\}$, $\{4,6\}$ and $\{5\}$, and those of $\tau$ are $\{1,3,6,7\}$, $\{4,5\}$ and $\{2\}$; 
and $\Asc\,\pi=\Atop\,\tau=\{5,6\}$.
}
\end{figure}

For a permutation $\pi\in \Ss_n$, a maximal interval $\{i,i+1,i+2,\dots,k\}\subset[n]$ is called {\it descent run} of $\pi$ if $\pi_i\pi_{i+1}\dots\pi_k$ is a decreasing factor of $\pi$, and if an entry $\pi_\ell$ does not belong to such a decreasing factor, then $\{\ell\}$ is a singleton descent run.
A subset $\{i_1,i_2,\dots,i_k\}$ of $[n]$ is called {\it inverse descent run}  (or {\it i.d.r.} for short) if $\{\pi_{i_1},\pi_{i_2},\dots,\pi_{i_k}\}$ is a descent run in the (group theoretical) inverse $\pi^{-1}$ of $\pi$. 
Alternatively, an i.d.r. can be defined  as a maximal (possibly singleton) subset $\{i_1,i_2,\dots,i_k\}$ of $[n]$ with $\pi_{i_\ell}=\pi_{i_{\ell+1}}+1$, for $1\leq \ell<k$. 
For a permutation $\pi$ in $\Ss_n$ with $k$ i.d.r's let $I_1,I_2,\dots,I_k$ be its i.d.r's. Then $\{I_1,I_2,\dots,I_k\}$ is a partition of $[n]$
and $\{\pi_i\}_{i\in I_\ell}$ is an interval for every $\ell$, $1\leq \ell\leq k$. In the following we order $I_1,I_2,\dots, I_k$ decreasingly by their largest element, that is, such that $\max(I_1)>\max(I_2>)\dots>\max(I_k)$. 
For instance, the i.d.r.'s of $\pi=7615324$ are $\{1,2,4,7\}$, $\{5,6\}$ and $\{3\}$, and those of $\tau=7132654$ are $\{1,5,6,7\}$, $\{3,4\}$ and $\{2\}$. See Figure \ref{fig_theta} where permutations are represented by their diagrams.

With the notations above, $\pi_i>\pi_j$ for every $i\in I_\ell$ and $j\in I_{\ell+1}$ and 
the following remark gives some easy to understand properties of i.d.r.'s of permutations in $\Av(231)$.

\begin{Rem}
For $\pi\in\Av(231)$ we have:
\begin{itemize}
\item $I_1$ is the set of positions of right-to-left maxima of $\pi$. Moreover, if we erase the entries $\pi_i$ in $\pi$ for each $i\in I_1$, then $I_2$ is the set of positions of right-to-left maxima of the resulting permutation, and so on;
\item $\{\max(I_1),\max(I_2),\dots,\max(I_k)\}$ is the set of positions of right-to-left minima of $\pi$.
\end{itemize}
\end{Rem}

For two finite sets $A$ and $B$ of integers we say that $B$ is {\it nested} in $A$ if there are no three integers $i<\ell<j$ with 
$i,j\in B$ and $\ell\in A$. 

\medskip
\noindent
The next proposition gives the characterization of permutations in $\Av(231)$ in terms of i.d.r.'s.

\begin{Pro}
The permutation $\pi$ belongs to $\Av(231)$ if and only if for every two i.d.r.'s $I_a$ and $I_b$ of $\pi$ with $\max(I_a)>\max(I_b)$ we have that $I_b$ is nested in $I_a$.
\end{Pro}
\proof
Clearly, if $I_b$ is nested in $I_a$ whenever $\max(I_a)>\max(I_b)$, then $\pi$ avoids $231$.
Conversely, if $I_b$ is not nested in $I_a$ with $\max(I_a)>\max(I_b)$, then there are 
$i,j\in I_b$ and $\ell\in I_a$ with $i<\ell<j$, and so $\pi_i\pi_\ell\pi_j$ is an occurrence of $231$ in $\pi$.
\endproof

The following technical definition will be used in the next two propositions.
\begin{De}
For $1\leq k\leq n$ the pair of sequences $c_1,c_2,\dots,c_k$ and $m_1,m_2,\dots,m_k$ of positive integers is {\it consistent} (with respect to $k$ and $n$) if:
\begin{itemize}
\item[] $c_1\geq 2$,
\item[] $c_1+c_2+\dots +c_k=n$,
\item[] $n=m_1>m_2>\dots >m_k$, and
\item[] $m_\ell\geq c_\ell+c_{\ell+1}+\dots+c_k+1$ for every $\ell$, $1<\ell\leq k$.
\end{itemize}
\end{De}

For example 
$c_1,c_2,c_3=4,2,1$ and $m_1,m_2,m_3=7,6,5$ is a pair of consistent sequences with $k=3$ and $n=7$.

\medskip
\noindent
We denote by $\Av'_n(231)$ the set of permutations in $\Av_n(231)$ beginning by $n$, and $\Av'(231)=\cup_{n\geq 0}\Av'_n(231)$.

\begin{Pro}
For $1\leq k\leq n$ let $c_1,c_2,\dots,c_k$ and $m_1,m_2,\dots,m_k$ be a pair of consistent sequences.
Then there is a unique permutation $\pi$ in $\Av'_n(231)$, where $n=m_1$, having $k$ i.d.r.'s, say $I_1,I_2,\dots,I_k$ ordered by $\max(I_1)>\max(I_2)>\dots>\max(I_k)$,  such that
\begin{enumerate}
\item $|I_\ell|=c_\ell$, $1\leq \ell\leq k$, and
\item $\Asc\,\pi=\{m_k,m_{k-1},\dots,m_2\}$.
\end{enumerate}
\label{car_maj_makl_Asc}
\end{Pro}
\proof 
We proceed by induction on $k$, and when $k=1$ the desired permutation $\pi$ is the 
decreasing one. 
For $k\geq 2$, the pair of sequences $c_1,c_2,\dots,c_{k-1}$ and $m_1-c_k,m_2-c_k,\dots,m_{k-1}-c_k$ is consistent with respect to 
$k-1$ and $n-c_k$, and let $\sigma$ be the permutation in $\Av'(231)$ of length $c_1+c_2+\dots c_{k-1}=n-c_k$ 
with $k-1$ i.d.r.'s, $I'_1,I'_2,\dots,I'_{k-1}$, and

\vspace{-0.29cm}
\begin{itemize}
\item $|I'_\ell|=c_\ell$, $1\leq \ell\leq k-1$, and
\vspace{-0.29cm}
\item $\Asc\,\sigma=$
$\{m_{k-1}-c_k,m_{k-2}-c_k,\dots,m_2-c_k\}$.
\end{itemize}
\vspace{-0.26cm}
The desired permutation $\pi$ is obtained from $\sigma$ by
adding $c_k$ to each element of $\sigma$, then inserting a contiguous i.d.r. $c_k(c_k-1)\cdots 1$ into the slot after its $(m_k-c_k)$th entry.
Formally, the permutation $\pi$ is defined as
$$
\pi_i=
\left\{ \begin {array}{ccl}
\sigma_i + c_k& {\rm if} & 1\leq i \leq m_k-c_k \mbox{ or } m_k < i\leq n \\
m_k-i+1 & {\rm if} & m_k-c_k+1\leq i\leq m_k.
\end {array}
\right.
$$
The permutation $\pi$ has $k$ i.d.r.'s $I_1,I_2,\dots,I_k$, and after deleting each entry $\pi_i$, $i\in I_k$, in $\pi$ and re-scaling the remaining entries to a permutation we recover $\sigma$.
It is routine to check that $\pi$ is the unique permutation satisfying conditions 1 and 2 in the present proposition, and the statement
holds. 
\endproof

\begin{Exam}
For $k=3$ and $n=7$, using the example of a
pair of consistent sequences $c_1,c_2,c_3=4,2,1$ and $m_1,m_2,m_3=7,6,5$ already given, the unique  permutation prescribed by Proposition \ref{car_maj_makl_Asc} 
is $\pi=7653124\in\Av_7'(231)$ with i.d.r.'s $I_1=\{1,2,3,7\}$, $I_2=\{4,6\}$, $I_3=\{5\}$ and $\Asc\,\pi =\{5,6\}$. See also Table~\ref{Table-3}, and Figure~\ref{fig_theta} for the diagram representation of $\pi$.
\end{Exam}

\begin{Pro}
For $1\leq k\leq n$ let $c_1,c_2,\dots,c_k$ and $m_1,m_2,\dots,m_k$ be a pair of consistent sequences.
Then there is a unique permutation $\tau$ in $\Av'_n(231)$, where $n=m_1$, having $k$ i.d.r.'s, say $I_1,I_2,\dots,I_k$ ordered by $\max(I_1)>\max(I_2)>\dots>\max(I_k)$,  such that
\begin{enumerate}
\item $|I_\ell|=c_\ell$, $1\leq \ell\leq k$, and
\item $\Atop\,\tau=\{m_k,m_{k-1},\dots,m_2\}$.
\end{enumerate}
\label{car_maj_makl_Atop}
\end{Pro}
\proof
Reasoning by induction as in the proof of Proposition \ref{car_maj_makl_Asc}, when $k=1$ the desired permutation $\tau$ is the decreasing one. 
For $k\geq 2$, let $j\leq k$ be the smallest integer such that the sets 
$1+c_{j+1}+c_{j+2}+\dots +c_k,2+c_{j+1}+c_{j+2}+\dots +c_k,\dots, c_j+c_{j+1}+c_{j+2}+\dots+c_k$
and $\{m_2,\dots,m_k\}$ are disjoint. Such a $j$ necessarily exists since 
$\{1,2,\dots,c_k\}$ and $\{m_2,\dots,m_k\}$ are disjoint.
Now, if $m_p$ is the smallest element of $m_1,m_2,\dots,m_k$ larger than $n-(c_1+c_2+\dots+ c_{j-1})$,
we denote by $m'_1,m'_2,\dots,m'_{k-1}$ the sequence $m_1-c_j,m_2-c_j,\dots,m_{p-1}-c_j,m_{p+1},\dots,m_k$ and by $c'_1,c'_2,\dots, c'_{k-1}$ the sequence  $c_1,c_2,\dots,c_{j-1},c_{j+1},\dots,c_k$.
The obtained pair of sequences $c'_1,c'_2,\dots,c'_{k-1}$ and $m'_1m'_2,\dots,m'_{k-1}$ is consistent with respect to 
$k-1$ and $n-c_j$ and let $\sigma$ be the permutation with $k-1$ i.d.r.'s, $I'_1,I'_2,\dots,I'_{k-1}$, and
\vspace{-0.29cm}
\begin{itemize}
\item $|I'_\ell|=c'_\ell$, $1\leq \ell\leq k-1$, and
\vspace{-0.29cm}
\item $\Atop\,\sigma=\{m'_{k-1},m'_{k-2},\dots,m'_2\}$.
\end{itemize}
\vspace{-0.26cm}
\noindent
The desired permutation $\tau$ is obtained from $\sigma$ by adding $c_j$ to each entry of $\sigma$ larger than 
$c_{j+1}+c_{j+2}+\dots+c_k$, then inserting a contiguous i.d.r. 
$(c_j+c_{j+1}+c_{j+2}+\dots+c_k)(c_j+c_{j+1}+c_{j+2}+\dots+c_k-1)\cdots (c_{j+1}+c_{j+2}+\dots+c_k+1)$ 
of length $c_j$
in the slot before the entry of value $m_p$.
The obtained permutation $\tau$ has $k$ i.d.r.'s, namely $I_1,I_2,\dots,I_k$, and after deleting each entry $\tau_i$, $i\in I_j$, in $\tau$ and re-scaling the remaining entries to a permutation
we obtain $\sigma$.
It is routine to check that $\tau$ is the unique permutation satisfying conditions 1 and 2 in the present proposition, and the statement holds. 
\endproof

\begin{Exam}
For $k=3$ and $n=7$, using the example of a
pair of consistent sequences $c_1,c_2,c_3=4,2,1$ and $m_1,m_2,m_3=7,6,5$, the unique  permutation prescribed by Proposition \ref{car_maj_makl_Atop}
is $\tau=7163254\in\Av_7'(231)$ with i.d.r.'s $I_1=\{1,3,6,7\}$, $I_2=\{4,5\}$, $I_3=\{2\}$ and $\Atop\,\tau=\{5,6\}$.
See also Table~\ref{Table-3}, and Figure~\ref{fig_theta} for the diagram representation of $\tau$.
\end{Exam}

\begin{table}[h]
\begin{center}
\begin{tabular}{|cc|c|}
\hline
\multicolumn{2}{|c|}{sequences} &  $\pi$ corresponding \\ 
$c$&$m$&  by Pr. 
2\\
\hline\hline
$4,2,1$ & $7,6,5$ & $7653\mathbf{1}24$\\
\hline 
$4,2$ & $6,5$& $654\mathbf{21}3$\\
\hline
4&4& $4321$\\
\hline
\end{tabular}
\begin{tabular}{|cc|c|}
\hline
\multicolumn{2}{|c|}{sequences} &  $\tau$ corresponding\\ 
$c$&$m$&  by Pr. 
3\\
\hline\hline
$4,2,1$ & $7,6,5$ & $716\mathbf{32}54$\\
\hline 
$4,1$ & $5,4$ & $5\mathbf{1}432$\\
\hline
4&4& $4321$ \\
\hline
\end{tabular}

\caption{Correspondences in Propositions \ref{car_maj_makl_Asc}
and 
\ref{car_maj_makl_Atop}
between consistent pairs of sequences and permutations. Permutations are constructed 
by inserting i.d.r.'s, in bold. The image of $\pi$ through the bijection $\theta'$ defined in Theorem 
\ref{The_maj_makl} 
is $\tau$; and $\Asc\,\pi=\Atop\,\tau$.
\label{Table-3}}
\end{center}
\end{table}

\begin{The}
For every $n\geq 1$, there is a bijection $\theta'$ from $\Av'_n(231)$ into itself which 
\begin{enumerate}
\item preserves the number of i.d.r's and their respective cardinality, 
\item transforms $\Asc$ into $\Atop$, and 
\item preserves $\Rlmaxl$ and $\Rlminl$ statistics.
\end{enumerate}
\label{The_maj_makl}
\end{The}
\proof
Let $\pi$ be a permutation in $\Av'_n(231)$ with $k$ i.d.r's, say
$I_1,I_2,\dots,I_k$ ordered by $\max(I_1)>\max(I_2)>\dots>\max(I_k)$. Then $\pi$ 
determines the consistent (with respect to $k$ and $n$) pair $c_1,c_2,\dots,c_k$ and $m_1,m_2,\dots,m_k$ satisfying  points 1 and 2 in Proposition \ref{car_maj_makl_Asc}. In turn, this pair uniquely determines a permutation $\tau\in \Av'_n(231)$ satisfying points 1 and 2 in Proposition \ref{car_maj_makl_Atop}. 
Thus, $\tau$ has $k$ i.d.r's, say $I'_1,I'_2,\dots,I'_k$, ordered by $\max(I'_1)>\max(I'_2)>\dots>\max(I'_k)$ and $|I_\ell|=|I'_\ell|$, $1\leq \ell\leq k$. In addition $m_\ell\in\Asc\,\pi$ if and only if $m_\ell\in\Atop\,\tau$, $2\leq\ell\leq k$.

It follows that the transformation $\pi\mapsto\tau$ is a bijection satisfying points 1 and 2 in the present theorem. Moreover, 
since $c_1=|I_1|=|I'_1|$ we have 
\begin{align*}
\Rlmaxl\,\pi&=\{n-c_1+1,n-c_1+2,\dots, n\}\\
&=\Rlmaxl\,\tau, 
\end{align*}

\noindent 
and
\vspace{-0.75cm}

\begin{align*}\Rlminl\,\pi&=\{\pi_{\max(I_k)},\pi_{\max(I_{k-1})},\dots, \pi_{\max(I_1)}\}\\
&=\{n-c_1+1,n-c_1-c_2+1,\dots,n-c_1-c_2-\cdots-c_k+1\}\\
&=\{\tau_{\max(I'_k)},\tau_{\max(I'_{k-1})},\dots, \tau_{\max(I'_1)}\}\\
&=\Rlminl\,\tau,
\end{align*}

\noindent
and the statement holds.
\endproof

\noindent
See Table \ref{Table-3} for the construction of $\theta'(\pi)$ in the proof of the previous theorem and Figure \ref{fig_theta} for an example.


Since every $n$-length permutation $\pi$ satisfies
$\Des\,\pi=[n-1]\setminus \Asc\,\pi$, and in addition if $\pi_1=n$, then 
$\Dbot\,\pi=[n-1]\setminus\Atop\,\pi$, it follows that bijection $\theta'$ in the proof of
Theorem \ref{The_maj_makl} transforms $\Des$ statistic into $\Dbot$ one, and we have the next corollary.
\begin{Co}
The multistatistics $(\Des,\Rlmaxl,\Rlminl)$ and 
$(\Dbot,\Rlmaxl,\Rlminl)$ are equidistributed over the set $\Av'_n(231)$ of $n$-length $231$-avoiding permutations beginning by $n$.
\label{Co_on_av'}
\end{Co}

Any permutation $\pi\in\Av_n(231)$ can uniquely be written as a direct sum 
\begin{equation}
\pi=\pi^{(1)}\oplus \pi^{(2)}\oplus\cdots \oplus\pi^{(k)}
\label{direct_sum}
\end{equation}
 of permutations in $\Av'_n(231)$, for some $k\geq 1$, and we extend $\theta'$ defined in Theorem \ref{The_maj_makl} to $\theta$ on $\Av_n(231)$ as

\begin{equation}
\theta(\pi^{(1)}\oplus \pi^{(2)}\oplus\cdots\oplus\pi^{(k)})=
\theta'(\pi^{(1)})\oplus \theta'(\pi^{(2)})\oplus\cdots \oplus \theta'(\pi^{(k)}),
\label{def_psi}
\end{equation}
and clearly its inverse is 
$$
\theta^{-1}(\pi^{(1)}\oplus \pi^{(2)}\oplus\cdots\oplus\pi^{(k)})=
\theta'^{-1}(\pi^{(1)})\oplus \theta'^{-1}(\pi^{(2)})\oplus\cdots \oplus \theta'^{-1}(\pi^{(k)}),
$$
and since $\theta'$ is bijective, so is $\theta$.

For $\pi\in \Av_n(231)$ as in relation (\ref{direct_sum}), the set $\Lrmax\, \pi$ is determined by the sequence $|\pi^{(1)}|,|\pi^{(2)}|,\dots,|\pi^{(k)}|$ which is the same as $|\theta'(\pi^{(1)})|,|\theta'(\pi^{(2)})|,\dots,|\theta'(\pi^{(k)}|$ (see relation (\ref{def_psi})), and so $\theta$ preserves $\Lrmax$ statistic. 
In addition, $\theta$ (as $\theta'$) transforms $\Des$ statistic into $\Dbot$ one, and 
$\pi$ and $\theta(\pi)$ have the same number of i.d.r's with the same respective cardinality.
Thus $\theta$ preserves in addition $\Rlmaxl$ and  $\Rlminl$ and we have the next consequence of Corollary \ref{Co_on_av'}.

\begin{Pro}
The bijection $\theta$ on $\Av_n(231)$ transforms the multistatistic
$(\Des,\Lrmax,\Rlmaxl, \Rlminl)$ into $(\Dbot, \Lrmax,\Rlmaxl, \Rlminl)$ one.
\label{main_231}
\end{Pro}

For $\pi\in \Av(231)$ it is easy to see that $\makl\,\pi=\sum_{i\in \Dbot\,\pi}\,i$, and since $\maj\,\pi=\sum_{i\in \Des\,\pi}\,i$ we obtain the next result.

\begin{Co}
The statistics $\maj$ and $\makl$ are equidistributed over $\Av_n(231)$.
\label{maj_makl_231}
\end{Co}

The statistic $\maj$ has the same distribution over $\Av_n(132)$ and over $\Av_n(231)$, see \cite{DDJSS}, and we have the next consequence of Corollary \ref{maj_makl_231}.

\begin{Co}
The statistics $\maj$ over $\Av_n(132)$ has the same distribution as $\makl$ over $\Av_n(231)$.
\label{maj_makl_132_231}
\end{Co}

\subsection{The equidistribution of $\maj$ and $\makl$ over $\Av_n(312)$}

The complement-reverse transformation $\com\circ\rev:\Ss\mapsto\Ss$ maps a permutation in $\Av(231)$ into one in $\Av(312)$ and conversely, and in the following we will turn the equidistributions over $\Av_n(231)$ in Section \ref{first_312} into ones over $\Av_n(312)$.
The transformation $\com\circ\rev$ is its own inverse, and it satisfies:
\begin{Property} For every permutation $\pi\in\Ss_n$ we have
\begin{enumerate}
\item $i\in\Des\,\pi$ iff $n-i\in\Des\,\com\circ\rev(\pi)$,
\item $i\in\Dbot\,\pi$ iff $n-i+1\in\Dtop\,\com\circ\rev(\pi)$,
\item $(i,j)\in\Lrmax\,\pi$ iff $(n-i+1,n-j+1)\in \Rlmin\,\com\circ\rev(\pi)$,
\item $i\in\Rlmaxl\,\pi$ iff $n-i+1\in \Lrminl\,\com\circ\rev(\pi)$,
\item $i\in\Rlminl\,\pi$ iff $n-i+1\in \Lrmaxl\,\com\circ\rev(\pi)$.
\end{enumerate}
\label{Property}
\end{Property}

\noindent
In the next proposition, which is the $312$-avoiding counterpart of Proposition \ref{main_231}, we will make use of the following notations
\begin{itemize}
    \item $n-\Des\,\pi=\{n-i\,:\,i\in\Des\,\pi\}$,
    \item $n-\Dbot\,\pi=\{n-i\,:\,i\in\Dbot\,\pi\}$,
    \item $\Dtop\,\pi-1=\{i-1\,:\,i\in\Dtop\,\pi\}$.
\end{itemize}

%
%

\begin{Pro}
The statistics $(\Des,\Rlmin,\Lrminl,\Lrmaxl)$ and $(\Dtop-1,\Rlmin,\Lrminl,\Lrmaxl)$ are equidistributed over the set $\Av_n(312)$.
\label{Av_312}
\end{Pro}
\proof
The map $\pi\mapsto \sigma=\com\circ\rev\,(\theta ( \com\circ\rev (\pi)))$ is a bijection from $\Av_n(312)$ into itself, where $\theta$
is the bijection from $\Av_n(231)$ into itself defined in Section \ref{first_312}. 
For $\pi\in \Av_n(312)$ we have

$$
\begin{array}{rcll}
\Des\,\pi & = & n-\Des\, \com\circ\rev (\pi) & \mbox{(by point 1 of Property \ref{Property})}\\
 & = & n-\Dbot\, \theta ( \com\circ\rev (\pi)) & \mbox{(by Proposition \ref{main_231})}\\
           & = & \Dtop\, \com\circ\rev\,(\theta ( \com\circ\rev (\pi)))-1, &  \mbox{(by point 2 of Property \ref{Property})}
\end{array}
$$
and thus $\Des\,\pi=\Dtop\,\sigma-1$.  By points 3--5 of Property \ref{Property} and Proposition \ref{main_231} it is routine to check that
$(\Rlmin,\Lrminl,\Lrmaxl)\,\pi=
(\Rlmin,\Lrminl,\Lrmaxl)\,\sigma$, and the statement holds.
\endproof

\begin{Exam}
If $\pi=4675321\in\Av(312)$, then 
\vspace{-0.15cm}
\begin{itemize}
\item $\com\circ\rev (\pi)=7653124$,
\vspace{-0.2cm}
\item $\theta(\com\circ\rev (\pi))=\theta'(\com\circ\rev (\pi))=7163254$ (see Table \ref{Table-3} and Figure \ref{fig_theta}), and 
\vspace{-0.2cm}
\item $\com\circ\rev (\theta(\com\circ\rev (\pi)))=4365271\in\Av(312)$.
\end{itemize}
\vspace{-0.2cm}
Moreover, $\Des\,\pi=\Dtop\, \com\circ\rev\,(\theta ( \com\circ\rev (\pi)))-1=\{3,4,5,6\}$.
\end{Exam}

Since $\maj\,\pi=\sum_{i\in \Des\,\pi}\,i$, and for $\pi\in \Av(312)$ we have $\makl\,\pi=\sum_{i\in \Dtop\,\pi-1}\,i$ we obtain the next result.

\begin{Co}
The statistic $\maj$ and $\makl$ are equidistributed over $\Av_n(312)$.
\label{maj_makl_312}
\end{Co}

The statistic $\maj$ has the same distribution over $\Av_n(312)$ and over $\Av_n(213)$, see \cite{DDJSS}, and we have the next consequence of Corollary \ref{maj_makl_312}.

\begin{Co}
The statistic $\maj$ over $\Av_n(213)$ has the same distribution as $\makl$ over $\Av_n(312)$.
\label{maj_makl_213_312}
\end{Co}

\subsection{Some consequences}

Combining the results in the present section with other equidistributions proved in \cite{Amini} we obtain the following corollary. 

\begin{Co}
The statistic $\makl$ 

\begin{enumerate} 
\item[] over $\Av_n(231)$ has the same distribution as
\begin{itemize} 
\item[1.] $\mak$ over $\Av_n(132)$ or over $\Av_n(312)$,
\item[2.] $\bastp$ or $\foze$ over $\Av_n(132)$, and
\end{itemize}

\item[] over $\Av_n(312)$ has the same distribution as
\begin{itemize} 
\item[3.]  $\mak$ over $\Av_n(213)$ or over $\Av_n(231)$,
\item[4.] $\foze$ or $\bastpp$ over $\Av_n(231)$.
\end{itemize}
\end{enumerate}
\label{mak_makl}
\end{Co}

\noindent
From Corollaries \ref{maj_makl_231}--\ref{mak_makl} together with results in \cite{JoannaChen} we have the next consequence.

\begin{Co}
The statistic $\bast$ over
\begin{enumerate} 
\item $\Av_n(213)$ has the same distribution as $\makl$ over $\Av_n(231)$,
\item $\Av_n(231)$ has the same distribution as $\makl$ over $\Av_n(312)$,
\item $\Av_n(213)$ has the same distribution as $\mak$ over $\Av_n(132)$ or over $\Av_n(312)$,
\item $\Av_n(231)$ has the same distribution as $\mak$ over $\Av_n(213)$ or over $\Av_n(231)$,
\item $\Av_n(231)$ has the same distribution as $\bastpp$ or $\foze$ over $\Av_n(231)$.
\end{enumerate}
\label{bast_makl}
\end{Co}

\begin{table}
\begin{center}
\begin{tabular}{cc}
\begin{tabular}{|l|cc|}
  \hline
  \multirow{4}{*}{$\maj,\makl$} & $231,231$ & Cor. \ref{maj_makl_231}\\
   & $132,231$  & Cor. \ref{maj_makl_132_231}\\ 
   & $312,312$  & Cor. \ref{maj_makl_312}\\ 
   & $213,312$  & Cor. \ref{maj_makl_213_312}\\  \hline
  \multirow{4}{*}{$\mak,\makl$} & $132,231$ & \multirow{2}{*}{Cor. \ref{mak_makl}.1 }\\
  & $312,231$ & \\
  & $213,312$ &  \multirow{2}{*}{Cor. \ref{mak_makl}.3 }\\
  & $231,312$ & \\ \hline
$\bastp,\makl$  &  132,231         & Cor. \ref{mak_makl}.2 \\ \hline
$\bastpp,\makl$  &  231,312         & Cor. \ref{mak_makl}.4 \\ \hline
   \multirow{2}{*}{$\foze,\makl$} & $132,231$ & Cor. \ref{mak_makl}.2 \\
  & $231,312$ & Cor. \ref{mak_makl}.4\\
\hline
   \multirow{2}{*}{$\bast,\makl$} & $213,231$ & \multirow{2}{*}{Cor. \ref{bast_makl}.1-2 }\\
  & $231,312$ & \\
\hline
\end{tabular}
&
\begin{tabular}{|l|cc|}
\hline
  \multirow{4}{*}{$\mak,\bast$} & $132,213$ & \multirow{4}{*}{Cor. \ref{bast_makl}.3-4 }\\
  & $312,213$ & \\
  & $213,231$ & \\
  & $231,231$ & \\
\hline
$\bastpp,\bast$ & 231,231 & Cor. \ref{bast_makl}.5\\ \hline
$\foze,\bast$ & 231,231 & Cor. \ref{bast_makl}.5\\ \hline
\end{tabular}
\end{tabular}
\end{center}
\caption{The equidistributions conjectured in \cite{Amini} that are proved in Section 4.
\label{Tab_Summ_makl}
}
\end{table}

\noindent
{\bf Further research directions.}
It can be of interest to explore the techniques presented in this paper for other cases left open in \cite{Amini} or for permutations of a multiset.

\subsection*{Acknowledgement}

\noindent
This research was initiated when the authors visited the Vietnam Institute for Advanced Study in Mathematics (VIASM), Hanoi, in the academic year 2019-2020. The authors express their warmest thank to the VIASM for the hospitality and for the wonderful working environment. 
The research is partly funded by Vietnam National Foundation for Science and Technology Development (NAFOSTED) under grant number 102.01-2020.23. 

\noindent
The authors are grateful to the anonymous referees for their careful reading of the paper and 
for providing many useful suggestions.
Special thanks to Sergey Kirgizov for fruitful discussion on the topic.

\end{document}